\documentstyle[12pt, epsfig]{article}

\begin{document}

\begin{titlepage}
\title{\small\raggedleft{gr-qc/9712043}\\
\small\raggedleft{DAMTP Relativity Group Preprint: DAMTP-R97/56}\\
\small\raggedleft{Fermilab Preprint: Pub-97/364-A}\\
\center\LARGE Persistence Amplitudes from Numerical Quantum Gravity}
\author{\small P. D'Eath,\\
        \small DAMTP,\\
        \small Cambridge University,\\
        \small 3 Silver Street,\\
        \small Cambridge, CB3 9EW,\\
        \small UK \and \small A. Sornborger,\\
        \small NASA/Fermilab Astrophysics Group,\\
        \small Fermi National Accelerator Laboratory,\\
        \small Box 500,\\
        \small Batavia, IL 60510-0500\\
        \small USA}
\date{\small October 27, 1997}
\end{titlepage}

\maketitle

\begin{abstract}

The Euclidean quantum amplitude to go between data specified on an
initial and a final hypersurface may be approximated by the tree
amplitude  
\begin{equation}
exp(-I_{classical}/\hbar),
\end{equation}
where $I_{classical}$ is the Euclidean action of the classical
solution joining the initial and final data. In certain cases the
tree amplitude is exact. We study $I_{classical}$, and hence the
quantum amplitude, in the case of a spherically symmetric Riemannian
gravitational field coupled to a spherically symmetric scalar
field. The classical scalar field obeys an elliptic equation, which we
solve using relaxation techniques, in conjunction with the field
equations giving the gravitational field. An example of the transition
from linearity to non-linearity is presented and power law behavior of
the action is demonstrated.

PACS numbers 04.20.-q, 04.20.Jb, 04.60.+n

\end{abstract}

\section{Introduction}

Much can be learnt in quantum gravity by considering the amplitude to
go from an initial to a final configuration of a gravitational and
a massless scalar field $\phi$. To complete the specification of the
amplitude, one needs to give the time separation between the two
spacelike surfaces depicted, together with other data at spatial
infinity [1]. If, for example, the time separation is taken to be
large, and the bounding data are taken to be weak, then one has a
scattering configuration. On the the other hand, the formalism is 
equally well adapted to studying strong-field amplitudes, as we shall
see in this paper, with the help of numerical analysis. One such
example concerns the late-time evolution of a radiating black hole.
Further examples are provided in quantum cosmology [2-4], where one
again needs the strong-field description.

Formally (if the time separation is taken to be Lorentzian) the
amplitude is given by a Lorentzian path integral

\begin{equation}
amplitude = \int e^{(iS/\hbar)} \label{onepointone}.
\end{equation}

Here the fields $g_{\mu\nu}$ and $\phi$ in the integration must agree
with the boundary data, the intrinsic three-dimensional metric
$h_{ij}$ and the scalar field $\phi$.  Quite apart from convergence
questions, there is a more elementary difficulty with the Lorentzian
path integral.  One might expect the path integral (\ref{onepointone})
to admit a semi-classical expansion

\begin{equation}
amplitude \sim (A+\hbar A_1+\cdots)exp(iS_c/\hbar)
\label{onepointtwo}.
\end{equation}

Here $S_c$ denotes the classical action (if such exists) of a
classical solution joining the initial to the final data.  It is a
functional of the boundary data and data at infinity. Similarly the
one-loop term A, two-loop term $A_1$ and so forth depend on the
boundary data.  Now consider the special case in which the classical
geometry is (nearly) flat and one has a massless scalar field between
an initial surface $t_1=0$ where $\phi =0$ and a final surface at time
$t_2$ where $\phi$ is non-zero [4].  The classical solution is

\begin{equation}
\phi(t,{\bf x})=const. \int d^3k {\tilde \phi(t_2,{\bf k})\over 
sin(\vert{\bf k}\vert t_2)}e^{i{\bf k}.{\bf x}} sin(\vert {\bf k}\vert
t).   \label{onepointthree}
\end{equation}

For general final data, the `solution' is singular because of the
poles in the denominator.  This is an example of a well-known
phenomenon, that the boundary-value problem for a hyperbolic system of
equations is not well posed [5].  The expression (\ref{onepointthree})
does, however, make it clear that the boundary-value problem is
well-posed if the time-separation is rotated by a small amount
$i\epsilon$ into the complex plane[6].

The most natural arena for studying boundary-value problems is within
Riemannian geometry, where the Lorentzian time coordinate $t$ is
replaced by $t=-i\tau$, with a positive-definite four-metric
$g_{\mu\nu}$ and massless scalar field $\phi$ [7].  One might
conjecture that the classical boundary-value problem is well-posed for
moderate-sized boundary data; indeed, the numerical methods to be
described in Sec. 2 will test this conjecture.  The conjecture has been
shown analytically to hold for weak (but non-linear) gravitational
fields in the absence of a scalar field [8].  The full question of the
existence of the coupled nonlinear elliptic Einstein-scalar field
equations subject to boundary data is a major unresolved problem in
general relativity and partial differential equations.  As suggested
in the previous paragraph, one would expect to obtain a Lorentzian
amplitude by starting with a Euclidean time interval at infinity, and
then rotating the Euclidean time interval towards a Lorentzian time
interval.  The first treatment of the black-hole evaporation problem
by Euclidean methods was by Hartle and Hawking [9].  The original
Euclidean approach to quantum cosmology was also due to Hartle and
Hawking [3]. 

For a Riemannian classical geometry, the quantum amplitude will admit
the semi-classical expansion 

\begin{equation}
amplitude\sim (A+\hbar A_1+\cdots)exp(-I_c/\hbar).  \label{onepointfour}
\end{equation}

Here $I_c$ is the classical action for gravity coupled to a massless
scalar, and $A, A_1,\cdots$ are one-, two-, and higher-loop terms,
dependent on the boundary data.  Even at the one-loop level, for pure
gravity, the Riemannian amplitude is known to be divergent, with
divergence linear in the surface invariants [4,10]

\begin{equation}
\begin{array}{ccl}
  I_1&=&\int d^3x h^{1/2} R_{ij}K^{ij},\\
  &&\\
  I_2&=&\int d^3x h^{1/2} (trK)^3,\\
  &&\\
  I_3&=&\int d^3x h^{1/2} K_{ij}K^{ij}(trK),\\
  &&\\
  I_4&=&\int d^3x h^{1/2} K^i_jK^j_kK^k_i.\\
  &&\\
\end{array}
\end{equation}

Here $R_{ij}$ is the Ricci tensor formed from the three-dimensional
metric $h_{ij}$, and $K_{ij}$ is the second fundamental form
[11]. (Without boundaries, pure gravity is divergent at two loops and
beyond [12].)  When a massless scalar field is added, there are again
one-loop surface divergences.

All these divergences may be cured by including supersymmetry,
providing that one continues to work with surface boundary data given
by $h_{ij}$ in the purely gravitational case, and with $h_{ij},\phi$
in the gravity-scalar case.  From this point of view quantum
corrections are `soft' in supersymmetric theories. In fact, provided
one sets the appropriate supersymmetric partners to zero on the
boundaries, it turns out that the amplitude just has the tree form
$exp(-I_c/\hbar)$, where $I_c$ is the classical action of the original
bosonic fields. One may use directly the supergravity calculations of
[4]. Note that there was a mistake in the paper [13] which included an
earlier discussion of the ideas of [4].  This has been corrected in
[4] by the inclusion of the auxiliary fields of supergravity [14],
which are of course needed for a complete treatment of supergravity.
Another way of motivating this calculation is to understand the
action as being derived as an effective action from low-energy
superstring theory, in which case we are calculating tree-level
amplitudes.
  
The local boundary conditions used here have been stressed in part
because they appear naturally in specifying the path integral.
However it is also important to understand that scattering boundary
conditions, by contrast, may have very different properties.  One
obtains scattering boundary conditions by pushing the two spacelike
surfaces off to infinity.  Then one takes an infinite product of zero-
or one- occupancy harmonic oscillator states.  This implies that there
is a non-trivial path integral involved in moving between local and
scattering boundary conditions, so that it is not surprising that
quantum amplitudes with the two types of boundary conditions should be
very different. Indeed [15], with scattering boundary conditions the
supersymmetric gravity-scalar model is divergent at one loop.

Since the quantum amplitude for the (supersymmetrized) gravity-scalar
model is exactly semi-classical, one strategy for computing such
amplitudes is clear: set up boundary data and solve the coupled field
equations numerically.  Include the classical Euclidean action $I_c$
in the numerical calculation, and finally compute the amplitude
$exp(-I_c/\hbar)$.  The equations for the fields and an expression for
the action are presented in Sec. 2. In Sec. 3, we discuss the
numerical methods used to find solutions. In Sec. 4, we present the
results obtained computationally. And, in Sec. 5, we discuss the
results and present our conclusions.

\section{Spherical Riemannian Scalar-Gravity Solutions}

The Riemannian metric is taken for simplicity in the spherically
symmetric form

\begin{equation}
g_{\mu\nu} = diag(e^b,e^a,r^2,{r^2}{sin^2}\theta) ,
\label{twopointone}
\end{equation}

\noindent
in coordinates $(t,r,\theta,\phi)$, where $t$ is a Riemannian time
coordinate and

\begin{equation}
\begin{array}{cc}
a=a(t,r),&b=b(t,r). \label{twopointtwo}
\end{array}
\end{equation}

\noindent
Notice that for real values of $a$ and $b$ in the metric, the metric
is positive definite.

The scalar field is also taken to be spherically symmetric:

\begin{equation}
\phi=\phi(t,r).  \label{twopointthree}
\end{equation}

\noindent
And the Euclidean action has the form

\begin{equation}
\begin{array}{ccl}
I&=&-{1\over 16\pi}\int {d^4}x g^{1\over 2} R\\
\\
&&\quad{}+{1\over 2} \int {d^4}x g^{1\over 2} (\nabla \phi)^2\\
\\
&&\quad{}+\hbox{boundary contributions}.\\
\\
\label{twopointfour}
\end{array}
\end{equation}

Here $g = det(g_{\mu\nu})$ and R is the Ricci scalar of
$g_{\mu\nu}$. The boundary contributions will later be important and
are discussed in Eqs.(2.19-22).  The field equations may be written in
the form 

\begin{equation}
\begin{array}{rcl}
R_{\mu\nu} &=& 8 \pi \phi_{,\mu} \phi_{,\nu}\\
\\
\partial_\mu(g^{1\over 2} g^{\mu\nu} \phi_{,\nu})&=&0\\
\\
\end{array}
\end{equation}

Here, we have set $G = \hbar = c = 1$.

The Lorentzian version of this problem, with a spherically symmetric
scalar field in a Lorentzian spherically symmetric gravitational
field, has been studied extensively, leading to many interesting
results [16-22].

Explicitly, the Euclidean gravitational field equations are

\begin{equation}
\begin{array}{ccl}
R_{tt} &=& {1\over 4}e^{-a+b}a'b'-{1\over 2}\ddot a -{1\over 4}\dot a^2\\
&&\quad{}+{1\over 4}\dot a \dot b -{1\over 2} e^{-a+b}b''-{1\over
4}e^{-a+b}b'^2 -e^{-a+b}{b'\over r}\\
&&\\
&=& 8 \pi {\dot \phi}^2\\
\\
R_{tr}&=&{\dot a\over r}\\
&&\\
&=&8 \pi {\dot \phi} \phi'\\
\\
R_{rr}&=&{1\over 4}a' b' +{a'\over r}-{1\over 2}e^{-b+a}\ddot a -{1\over
4} e^{-b+a}{\dot a}^2\\
&&\quad{}+{1\over 4}e^{-b+a}{\dot a}{\dot b}-{1\over 2}b''
-{1\over 4}{b'}^2\\
&&\\
&=&8 \pi (\phi')^2\\
\\
R_{\theta\theta}&=&{1\over 2}e^{-a}a'r-{1\over 2}e^{-a}b'r-e^{-a}+1\\
&&\\
& =&0\\
\\
\end{array}
\end{equation}

The field equations can be simplified, to give (starting with the
scalar field equation):

\begin{equation}
{\ddot \phi}+e^{b-a}\phi''+{1\over 2}(\dot a -\dot b){\dot \phi}
+{e^{b-a}\over r} (1+e^a){\phi'} = 0,\label{twopointeleven}
\end{equation}

\begin{equation}
a'=-4\pi r (e^{a-b}{\dot \phi}^2-{\phi'}^2)+{(1-e^a)\over r}\\
\label{twopointtwelve}
\end{equation}
\begin{equation}
b'=-4\pi r (e^{a-b}{\dot \phi}^2-{\phi'}^2)-{(1-e^a)\over r}\\
\label{twopointthirteen}
\end{equation}
\begin{equation}
\dot a = 8\pi r {\dot \phi}\phi'\\
\label{twopointfourteen}
\end{equation}

\begin{equation}
\ddot a +e^{b-a}b''+{1\over 2}(\dot a -\dot b)\dot a
-e^{b-a}{(1-e^a)\over r}(b'+{2\over r})=8\pi ({\dot
\phi}^2+e^{b-a}{\phi'}^2). \label{twopointfifteen}
\end{equation}

One can see the general structure of the coupled Riemannian
Einstein-scalar equations from this system.  The scalar field obeys an
elliptic equation (\ref{twopointeleven}) which is determined if the
gravitational background is known.  Conversely,
Eqs. (\ref{twopointtwelve} - \ref{twopointfifteen}) determine $a$ and
$b$ if $\phi$ is known.  An effective nonlinearity is present in the
elliptic Eq. (\ref{twopointeleven}) due to the presence of $a$ and $b$
which involve $\phi$. Notice that Eq. (\ref{twopointfifteen}) is
equivalent to Eq. (\ref{twopointeleven}) given the constraints,
i.e. they are not independent equations.

As the simplest example of a boundary-value problem, one might wish to
solve the field equations inside a rectangular boundary, where $r$
runs from the axis $r=0$ to an outer boundary $r=r_{max}$, and $t$ runs
from an initial value $t_i$ to a final value $t_f$. One normally
expects that $\phi$ and the intrinsic 3-metric $h_{ij}$ would be
specified on the boundary [4], in which case one would have $a=a_i$
fixed on the initial surface, $a_f$ on the final surface, and $b=b_f$
on the outer surface.  Note also that, for regularity of the initial
3-metric on the initial surface, one should have $a\rightarrow 0$ as
$r\rightarrow 0$ on the initial surface.  Now consider
Eq. (\ref{twopointfourteen}).  By regularity on the axis, $\phi' = 0$
there.  Hence $\dot a =0$ along the axis.  Hence

\begin{equation}
a(t,r=0)=0.   \label{twopointsixteen}
\end{equation}

\noindent
This provides another boundary condition.

However, we are unable to solve this version of the boundary-value
problem, as the data on the boundaries are overdetermined, we cannot
simultaneously fix $\phi$ and $a$ and $b$. The correct
approach is to fix $\phi$ on the boundaries, as well as fixing $b$ on
the outer boundary (thus fixing the gauge freedom for $\dot b$), then
solve Eq. (\ref{twopointeleven}) by relaxation [23] and then to
compute $a$ and $b$ by integrating the constraint equations
(\ref{twopointtwelve}, \ref{twopointthirteen}) iteratively, while
checking that the full set (\ref{twopointeleven} -
\ref{twopointfifteen}) holds after the iteration. There is also one
leftover gauge degree of freedom, the value of $\dot b$, which we
choose to fix at $r = r_{max}$ by fixing $b$ at $r_{max}$.

The classical action for our system of equations resides on the
boundary.  It is

\begin{equation}
I_c = -{1\over 8 \pi}\int {d^3}x h^{1\over 2} trK
      +{1\over 2} \int {d^3}x h^{1\over 2} \phi {\partial \phi
      \over{\partial n}}
.\label{twopointnineteen}
\end{equation}

\noindent
The integral is taken over all bounding surfaces, with intrinsic
metric $h_{ij}$ and $h=det(h_{ij})$.  Here $K_{ij}$ is the Euclidean
second fundamental form [4], taken with respect to the outward normal,
and $trK=h^{ij} K_{ij}$.  For the upper surface $t=t_f$ (say), one
finds

\begin{equation}
I_f = -{1\over 4}\int dr r^2 e^{{(a-b)}/2} \dot a
      +{2 \pi} \int dr r^2 e^{{(a-b)}/2} \phi \dot \phi
,\label{twopointtwenty}
\end{equation}
        
\noindent
with a corresponding expression for $I_i$ in which the minus and plus
signs are reversed.  There is also a contribution from the outer
boundary at $r=r_f$, which is near $r=\infty$ in our case.  It is [4]

\begin{equation}
I_\infty=MT,       \label{twopointtwentyone}
\end{equation}

\noindent
where $M$ is the mass of the space-time and $T$ is the proper
Euclidean time-interval between the initial and final surfaces.  Hence
the total classical Euclidean action is

\begin{equation}
\begin{array}{ccl}
I_{total} &=&-{1\over 4}\int_{top} dr r^2 e^{{(a-b)}/2} \dot a
      +{2 \pi} \int dr r^2 e^{{(a-b)}/2} \phi \dot \phi\\
\\
&&\quad{}+{1\over 4}\int_{bottom} dr r^2 e^{{(a-b)}/2} \dot a
      -{2 \pi} \int dr r^2 e^{{(a-b)}/2} \phi \dot \phi\\
\\
&&\quad{}+MT.\\
\\
\end{array}
\end{equation}

\section{Numerical Methods}

We use the standard artificial time method for relaxing the elliptic
equation. This relaxation method uses the artifice of introducing a
diffusive term to the elliptic equation. The diffusive term gradually
relaxes to zero and one is left with a solution to the elliptic
operator of interest.

With this relaxation method, we solve the equation

\begin{equation}
\partial_{\tau}\phi + L(\phi,a,b,\dot a,\dot b) = 0
\label{twopointseventeen}
\end{equation}

\noindent
where $\tau$ is an artificial time coordinate, and L is the nonlinear
elliptic operator for which we desire a solution. We solve the
equation by discretising in the artificial time and also imaginary
time and space, then integrating the equation in time. For a linear
elliptic operator discretised in this manner, there is a constraint on
the allowed size of the integration timestep. For instance, for
solving the heat equation, the constraint is $\Delta t \leq (\Delta
x)^2/2$. For a nonlinear elliptic operator, the timestep must
typically be kept substantially smaller, such that the integration is
effectively linear. In our integrations, for a simulation volume of
0.8, we took $\Delta t = 0.00005$, typically. The small value of
timestep needed in our case is a measure of the non-linearity of our
problem.

After one integration of Eq. (\ref{twopointseventeen}) where $L$ is
given by (\ref{twopointeleven}), we use Eq. (\ref{twopointtwelve}) to
integrate $a$ radially outwards from the axis where $a=0$. For a
radius of one gridpoint out from the axis we are forced to approximate
Eq. (\ref{twopointtwelve}) in order to avoid problems from divergences
due to small values of $r$. At large $r$, the solution for $a$ will be
dominated by the non-$\phi$ part of Eq. (\ref{twopointtwelve}). The
solution to this will be a Schwarzschild solution, so that one will
have $a \sim 2M/r$ as $r \sim \infty$, where $M$ is the mass of the
space-time.  One also integrates Eq. (\ref{twopointthirteen}) to
derive values for $b$ inwards from the outer boundary to the axis.
Then one iterates this procedure until $\phi, a, b$ converge (if this
occurs), checks that the remaining equations (\ref{twopointfourteen},
\ref{twopointfifteen}) hold, and evaluates the action.

\section{Results}

We have run a number of integrations with simulation volume sides of
$0.8$ in units where $G = \hbar = c = 1$. $\phi(r)$ was set to

\begin{equation}
\phi= {d\over{8 \pi}} exp(-100 r^2),  \label{twopointeighteen}
\end{equation}

\noindent
(where $d$ is a parameter) on the initial and final time surfaces.
Thus, we have identical values of $\phi$ at opposite points on the
initial and final surfaces.  Note here that the outer boundary at
$r = r_{max} = 0.8$ is fairly close to $r = \infty$ in terms of the
asymptotic fields. This means that the action is simple to calculate
at the $r_{max}$ boundary.

We performed integrations on grids with $25^2$, $35^2$, $50^2$, $75^2$
and $100^2$ gridpoints. We also performed a few integrations on a
$200^2$ grid, however, these calculations were very time consuming. 

\begin{figure}
\centerline{\psfig{figure=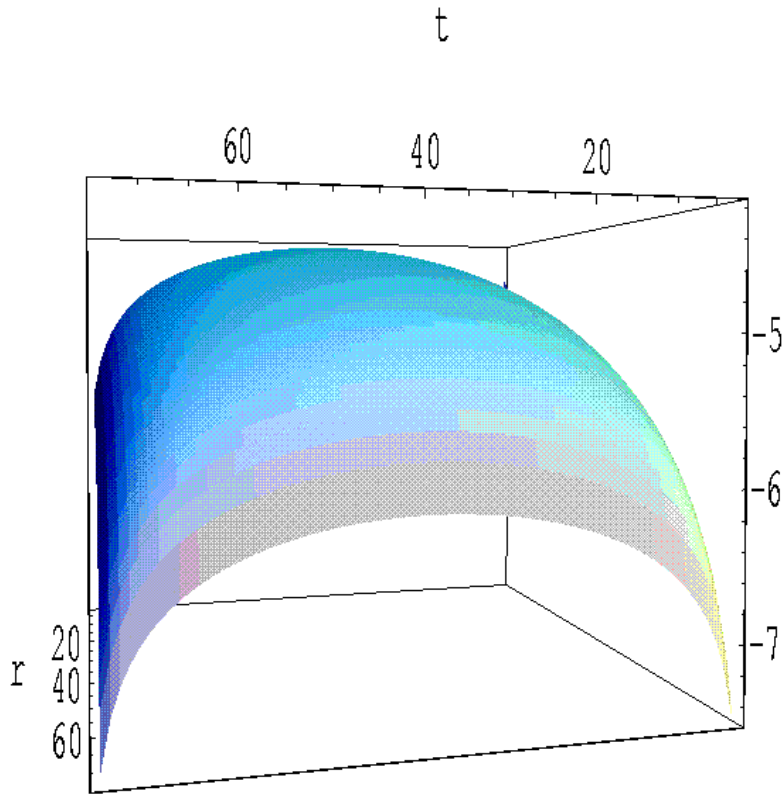,height=5.0in,width=5.0in}}
\caption{A plot of the $\log_{10}$ of the error in calculation of the
elliptic scalar field equation for $\phi$ on a $75^2$ grid for $d =
1.0$.
\label{straint75phi}}
\end{figure}

\begin{figure}
\centerline{\psfig{figure=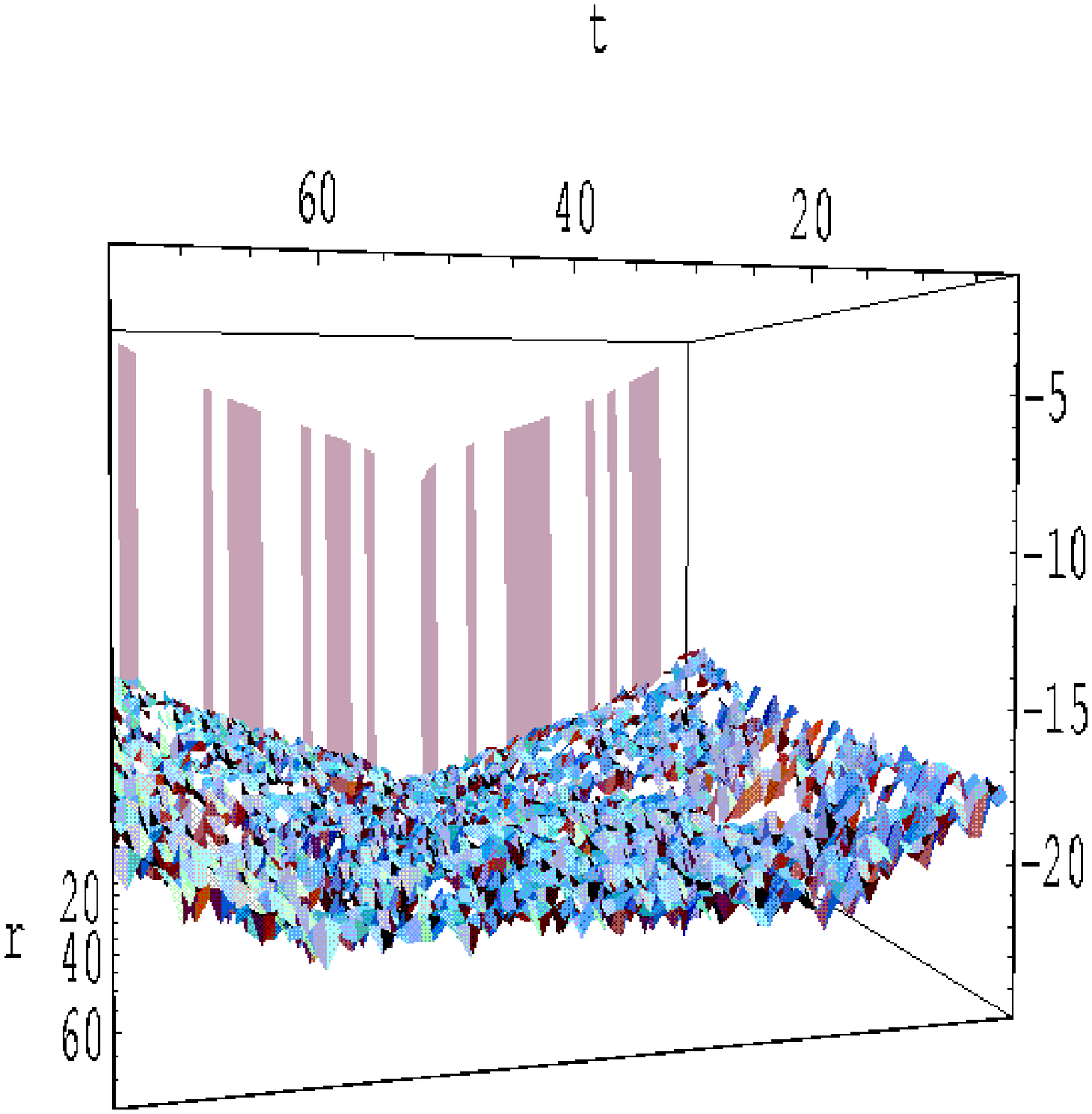,height=5.0in,width=5.0in}}
\caption{A plot of the $\log_{10}$ of the error in calculation of the
constraint equation for $a'$ on a $75^2$ grid for $d = 1.0$.
\label{straint75a}}
\end{figure}

\begin{figure}
\centerline{\psfig{figure=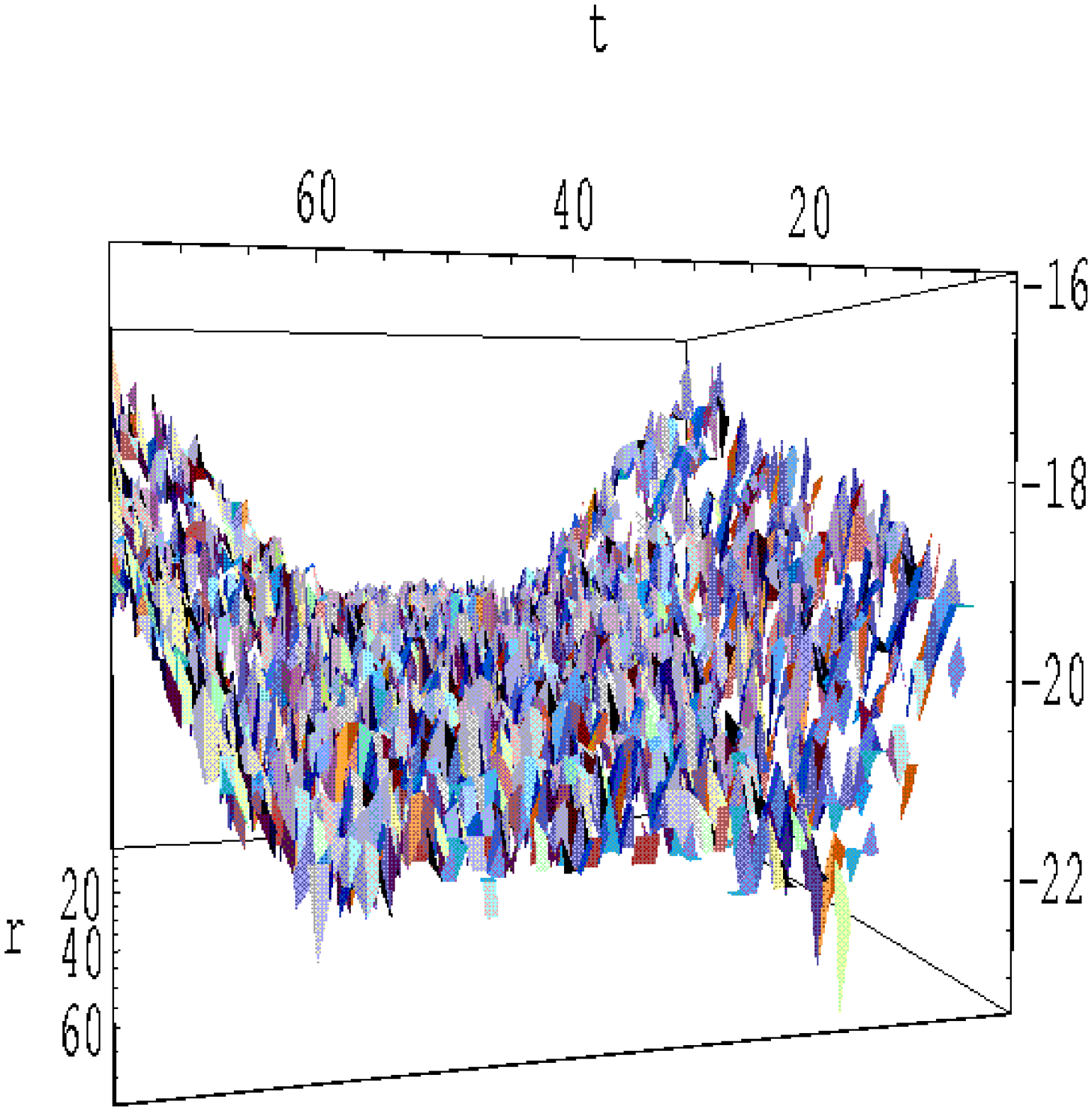,height=5.0in,width=5.0in}}
\caption{A plot of the $\log_{10}$ of the error in calculation of the
constraint equation for $b'$ on a $75^2$ grid for $d = 1.0$.
\label{straint75b}}
\end{figure}

\begin{figure}
\centerline{\psfig{figure=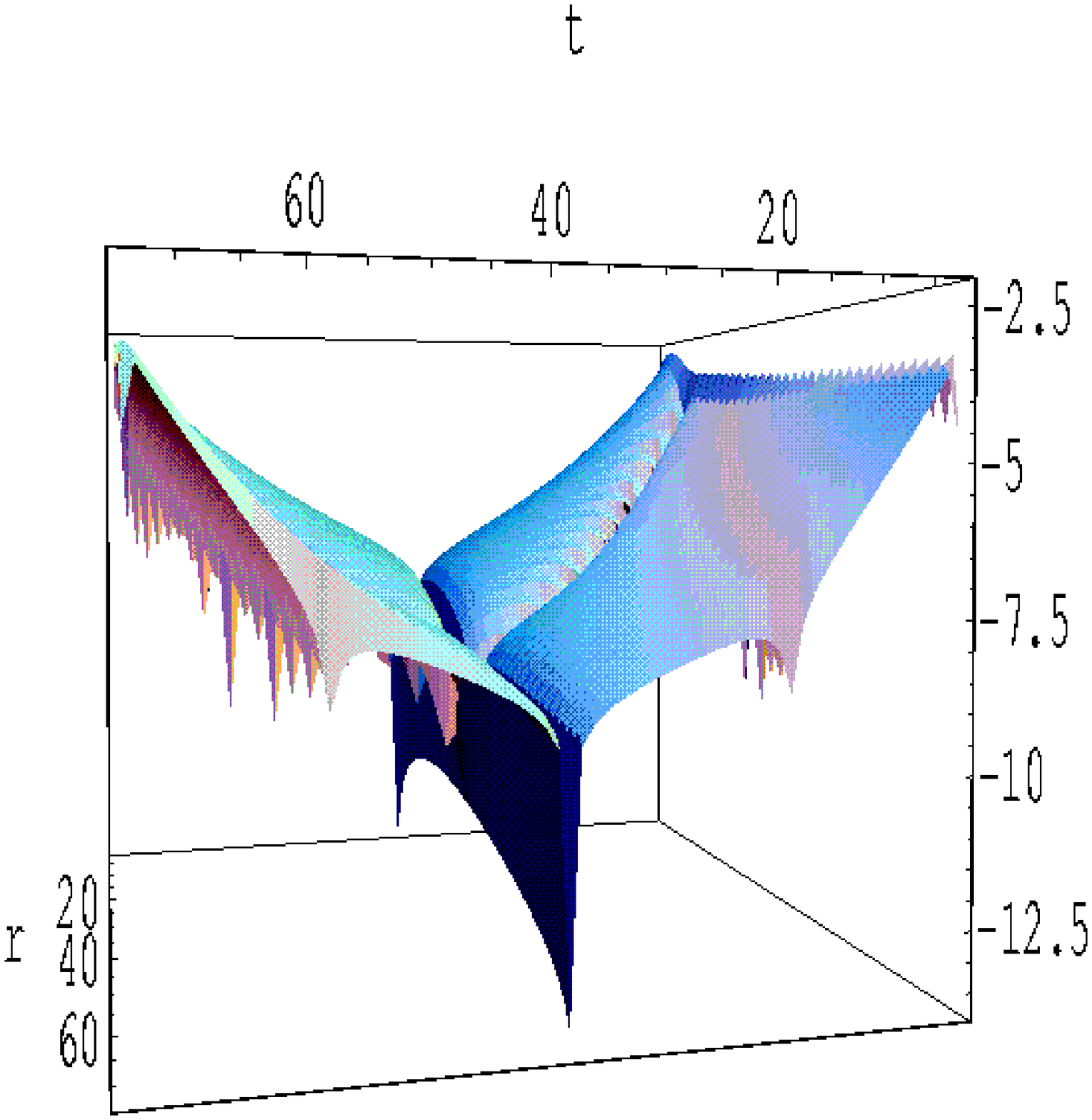,height=5.0in,width=5.0in}}
\caption{A plot of the $\log_{10}$ of the error in calculation of the
constraint equation for $\dot a$ on a $75^2$ grid for $d = 1.0$.
\label{straint75three}}
\end{figure}

To claim a solution, we required that the elliptical scalar field
equation be zero to one part in one hundred thousand. We also checked
that the constraint equations were satisfied. In general, the
elliptical equation converged to $10^{-5}$ up to some value of the
parameter $d$. Smaller grids tended not to converge for smaller values
of $d$ than larger grids. Because the constraint equations are just
ordinary differential equations integrated forwards or backwards in
the $r$ direction on the grid, they had roughly constant errors. In
figures (\ref{straint75phi}), (\ref{straint75a}), (\ref{straint75b})
and (\ref{straint75three}), we plot the $\log_{10}$ of the errors for
the solutions of equations (\ref{twopointeleven}),
(\ref{twopointtwelve}), (\ref{twopointthirteen}) and
(\ref{twopointfourteen}) for a solution on a $75^2$ mesh with $d =
1.0$. Note that the error for $\phi$ is everywhere less than
$10^{-5}$. The errors for the constraints on $a'$ and $b'$ are
negligible (parts of the plotted surface that are missing correspond
to errors of $0$ to within machine accuracy). And the error for the
constraint on $\dot a$ is less than $10^{-2.5}$. Significantly, the
error for the constraint equation for $\dot a$ is largest where
spurious oscillations occur in the solution (see below).

\begin{figure}
\centerline{\psfig{figure=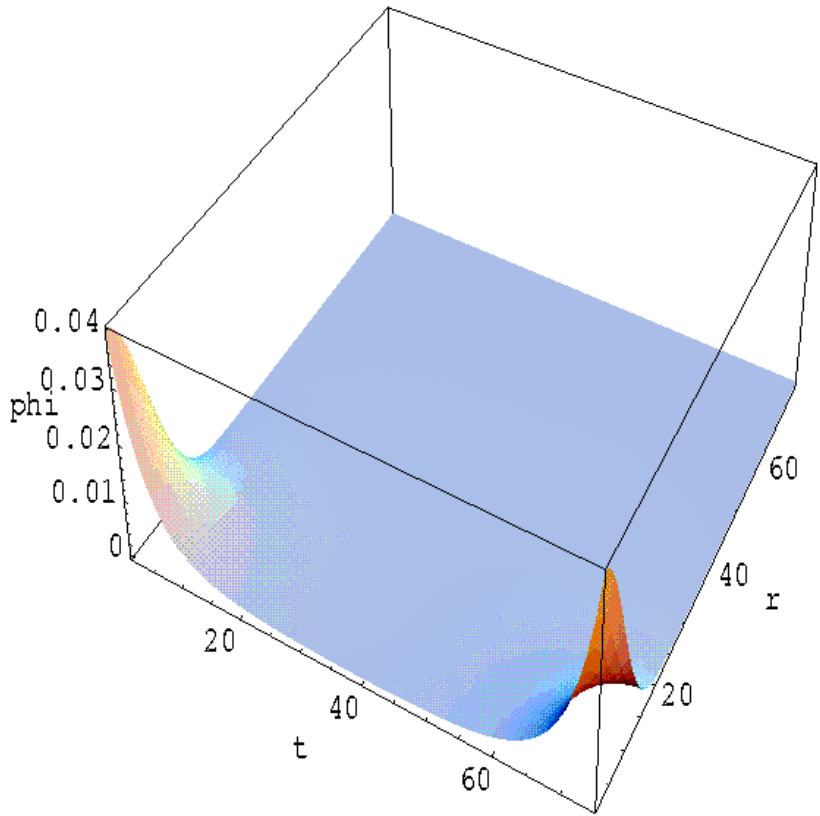,height=5.0in,width=5.0in}}
\caption{The solution for field $\phi$ on a $75^2$
grid. Notice the time symmetry of the solutions here and in figures
(\ref{dat75a}) and (\ref{dat75b}), this indicates that our numerical
methods are obeying the symmetry of the equations.
\label{dat75phi}}
\end{figure}

\begin{figure}
\centerline{\psfig{figure=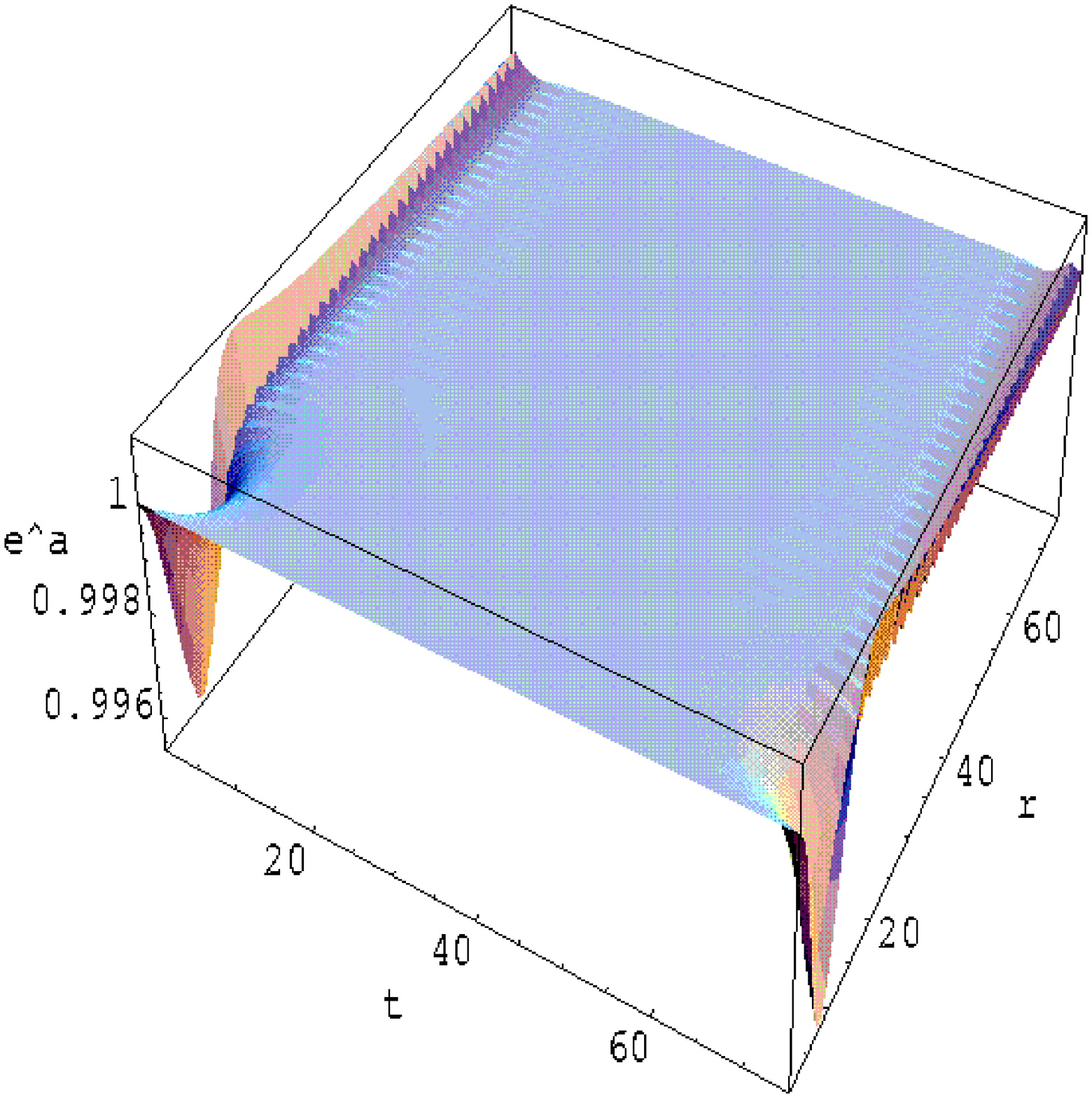,height=5.0in,width=5.0in}}
\caption{The solution for metric field $e^a$ on a $75^2$ grid
\label{dat75a}}
\end{figure}

\begin{figure}
\centerline{\psfig{figure=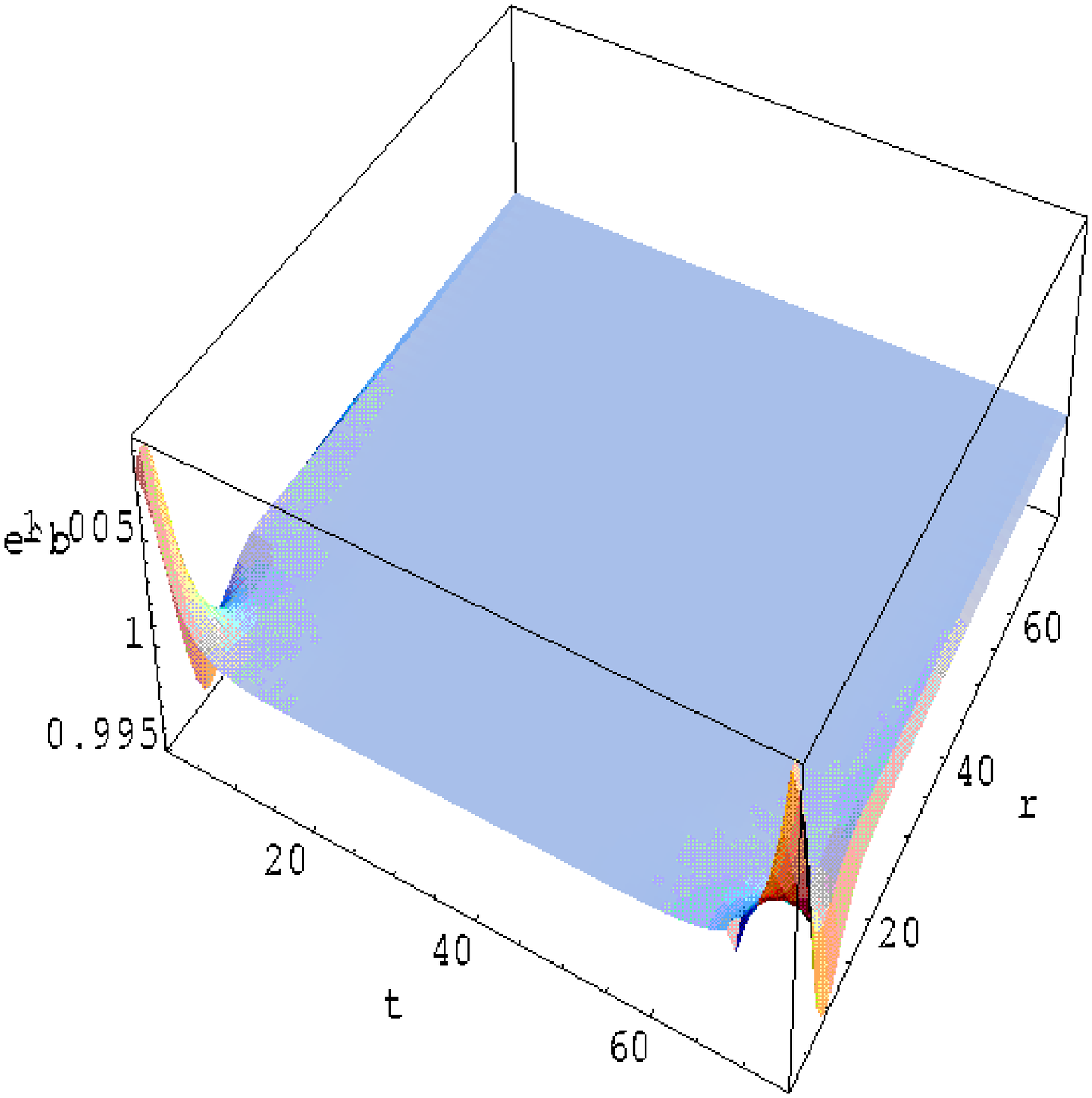,height=5.0in,width=5.0in}}
\caption{The solution for metric field $e^b$ on a $75^2$ grid
\label{dat75b}}
\end{figure}

In figures (\ref{dat75phi}), (\ref{dat75a}) and (\ref{dat75b}) the
solution for $d = 1.0$ is shown on a $75^2$ grid. Notice the time
symmetry in $\phi$ as well as $e^a$ and $e^b$. This is evidence that
our relaxation method preserves the time symmetry of the equations.

\begin{figure}
\centerline{\psfig{figure=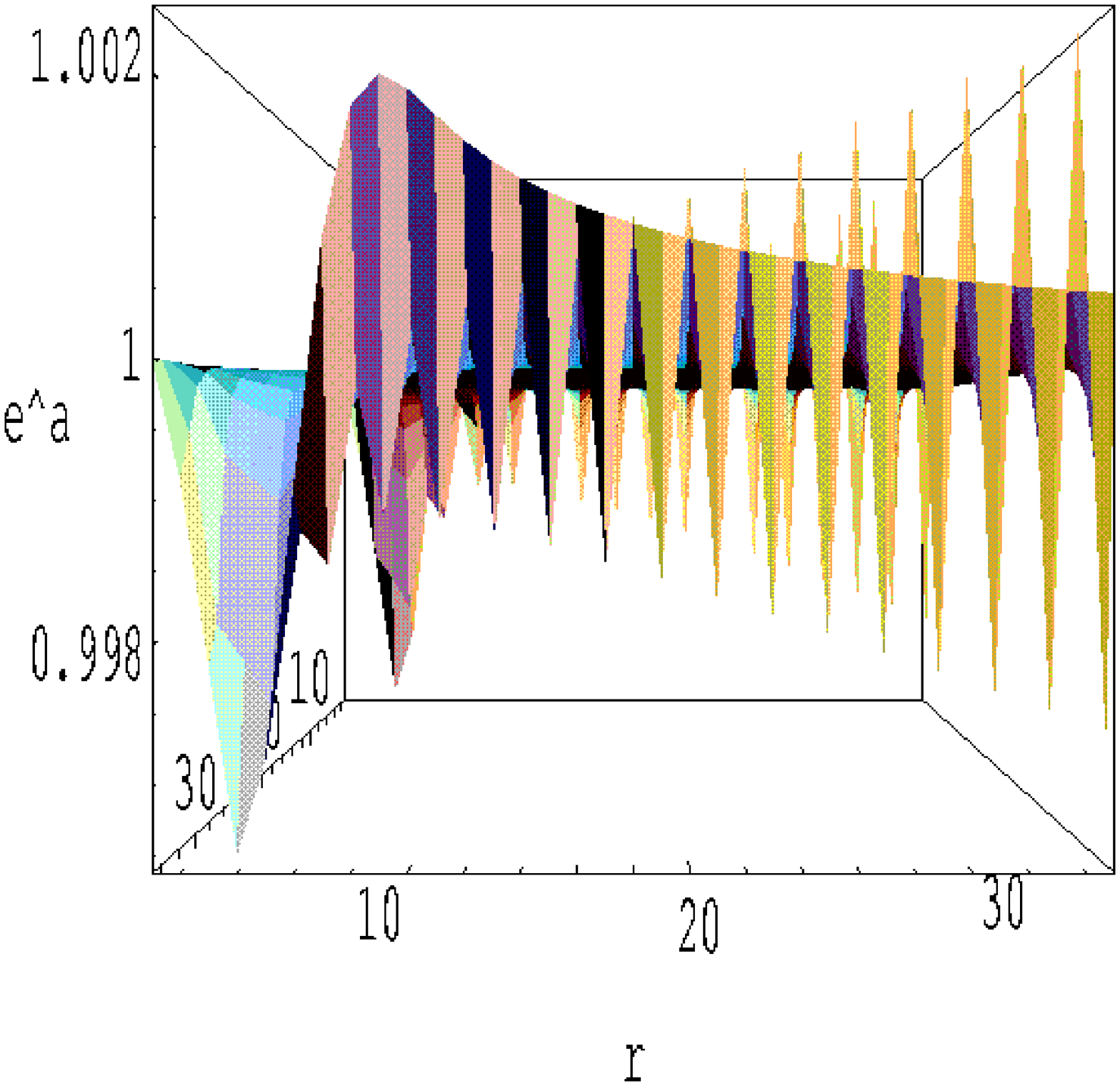,height=5.0in,width=5.0in}}
\caption{The solution for metric field $e^a$ on a $35^2$ grid. This is
a side view. The amplitude of the spurious oscillations on this coarse
grid decrease as the grid resolution increases in figures
(\ref{dat100acomp}) and (\ref{dat200acomp})
\label{dat35acomp}}
\end{figure}

\begin{figure}
\centerline{\psfig{figure=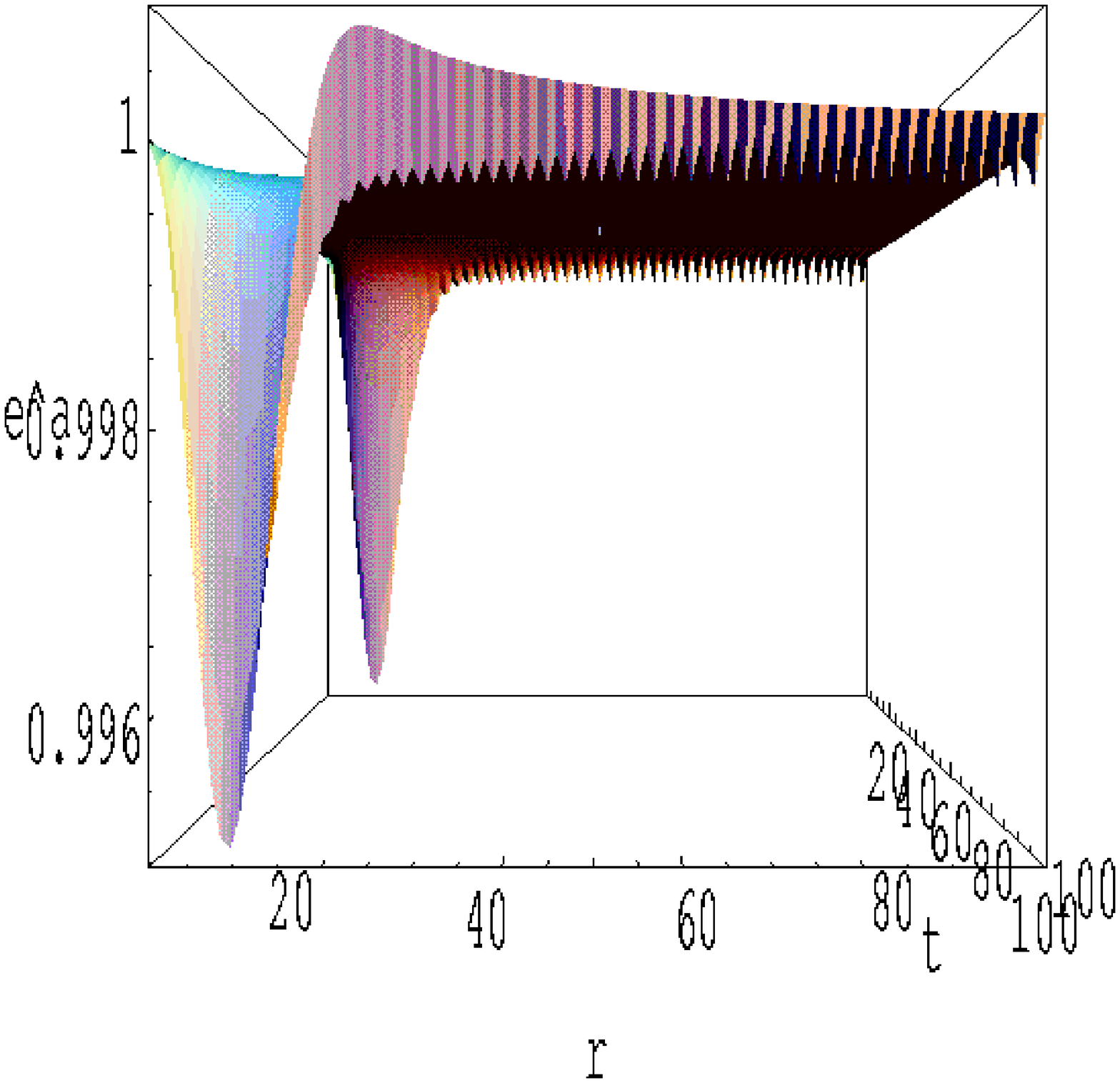,height=5.0in,width=5.0in}}
\caption{The solution for metric field $e^a$ on a $100^2$ grid.
\label{dat100acomp}}
\end{figure}

\begin{figure}
\centerline{\psfig{figure=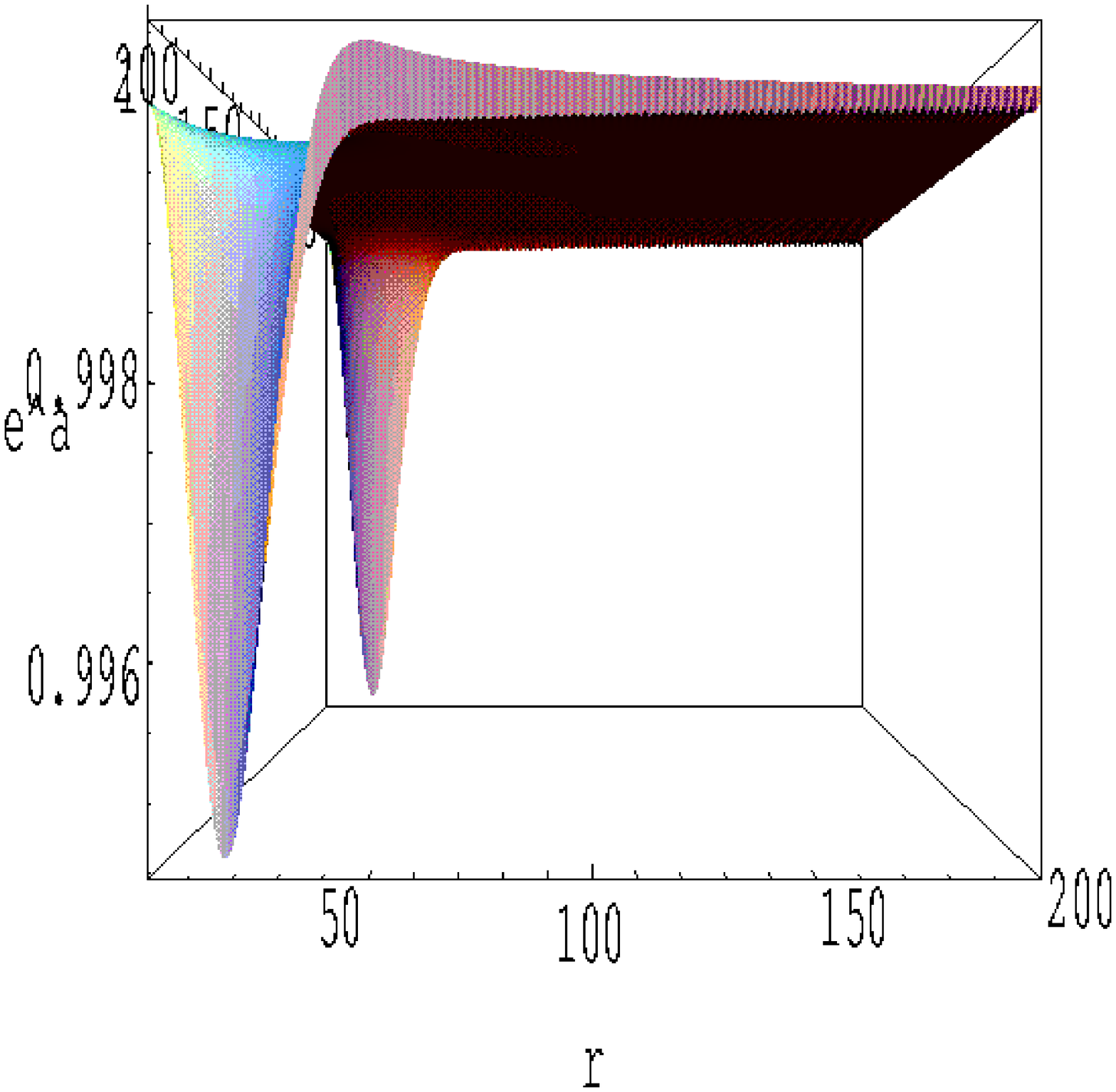,height=5.0in,width=5.0in}}
\caption{The solution for metric field $e^a$ on a $200^2$ grid.
\label{dat200acomp}}
\end{figure}

\begin{figure}
\centerline{\psfig{figure=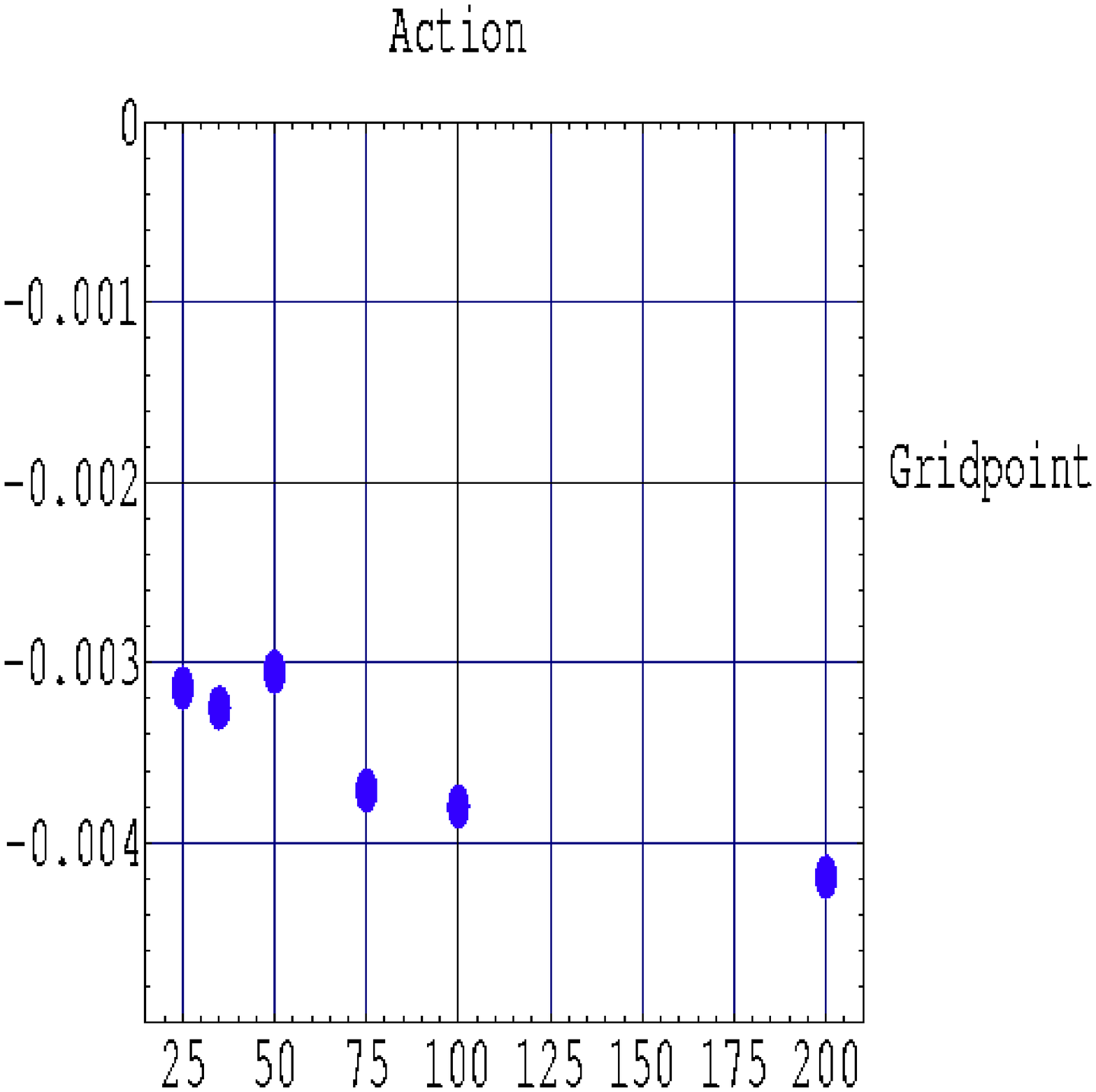,height=5.0in,width=5.0in}}
\caption{In this plot, we show the dependence of the value of the
action on the resolution of the numerical integration.
\label{actioncomp}}
\end{figure}

In all numerical solutions, we find spurious oscillations near, but
not at the boundary, as presented in figures (\ref{dat35acomp}),
(\ref{dat100acomp}) and (\ref{dat200acomp}). The oscillations occur
only in the $a$ metric component. Notice from the figures that the
amplitude of the oscillations decreases with finer meshes, with
oscillations on the $200^2$ grid substantially reduced with respect to
the coarser grids. Interestingly, the spurious oscillations do not
appreciably affect the value of the action. This is because the
oscillations are integrated over in calculating the action. A feature
that does affect the action calculation can be seen in the depth of
the dip near $r = 0$. This feature ranges in value from $0.9965$ for
the $35^2$ solution, to $0.9947$ for the $200^2$ solution. In figure
(\ref{actioncomp}), we see how this influences the value of the
action.

\begin{figure}
\centerline{\psfig{figure=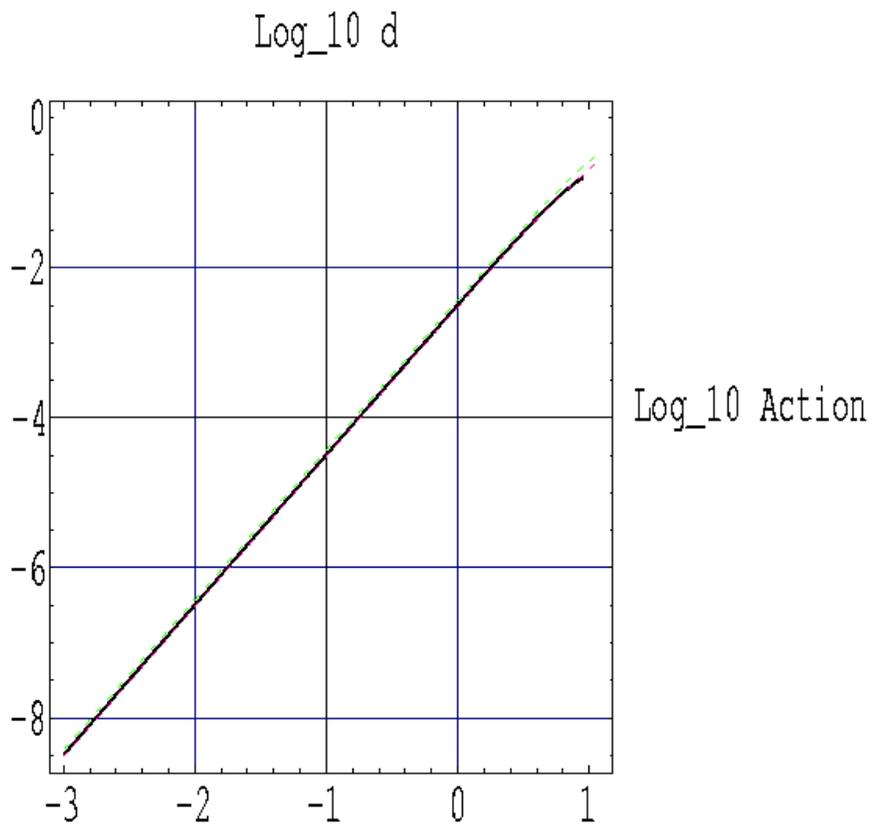,height=5.0in,width=5.0in}}
\caption{Here, we plot $\log_{10}$ of the absolute value of the action
as a function of $\log_{10}$ of parameter $d$ for gridsizes $35^2$,
$50^2$ and $100^2$. Thus, $I \sim -d^2$. This behavior shows the
expected increase of persistence amplitude with binding energy of the
system.
\label{powerlaw}}
\end{figure}

In figure (\ref{powerlaw}), we plot $\log_{10}$ of the action as a
function of $\log_{10}$ of the parameter $d$. The resulting power law
with slope $2$ is quite striking. In this figure, we have actually
plotted the absolute value of the action, which means that the action
$I \sim -d^2$. The decrease of the action with increasing $d$ is a
natural result of the increased binding energy of the system. As we
enter the nonlinear regime, we would expect that the power might
change, and, indeed, we do begin to see a bend for larger values of
$d$ in the plot. However, the bend decreases with increasing grid
resolution. For various grid sizes $L$, we find the slope $n$ to be:

\begin{equation}
\begin{array}{cccc}
&L^2&&n\\
&35^2&&1.956\\
&50^2&&1.968\\
&100^2&&1.971\\
\end{array}
\end{equation}

The straightening of the bend in the action is due to the better
resolution of the dip in the metric functions on higher resolution
meshes. The tendency for the slope to approach $2$ for high resolution
meshes, may indicate that the actual solution has a slope of $2$
exactly, even well into the nonlinear regime. However, we have only
shown the tendency, and cannot prove this assertion.

\section{Discussion and Conclusions}

In this paper, we have studied amplitudes for the spherically
symmetric system of a scalar field coupled to gravity. We have looked
at amplitudes for the field to persist in its current state. We have
taken the field to have support primarily near the symmetry axis, with
amplitude parametrised by the parameter $d$.

What we can say from our study, is that the Euclidean action is
well-behaved, negative and obeys obvious power law behaviour, with the
action $I \sim -d^2$. There are indications that the power behavior
remains the same into the nonlinear regime. The decrease of the action
with increasing $d$ is a result of the increasing binding energy of
the system as the scalar field gets large. This result is also
familiar from the understanding of black hole evaporation: massive
black holes radiate more than less massive black holes. The
persistence amplitudes that we calculate show the opposite side of the
coin: a massive scalar field is more likely to retain its
integrity than a less massive scalar field.

The solutions that we obtain are all time symmetric and largely slowly
varying (for values of $d$ which we have been able to probe) outside
of regions near the initial and final times, where they vary rapidly.

From the numerical solutions that we have obtained, it is clear that
more work needs to be done exploring the strong-field Euclidean
gravity equations. The obvious direction to go numerically is to
develop a mesh-refinement code. A mesh-refinement code would increase
the efficiency of the calculation by focusing on regions of the
solution that are varying quickly, thus decreasing the computational
time needed for a given resolution, and increasing the resolution of
the solution.

This conclusion from our work on fixed meshes is in accord with what
we know of critical behavior from work on the Choptuik
problem. Adaptive meshes with an effective resolution of billions of
gridpoints were necessary to see the full nonlinear critical behavior
in the Lorentzian case.

As well as refining the numerical approach to this problem, we plan to
investigate more different kinds of solution. In particular, solutions
which are not time symmetric, in order to investigate amplitudes for
evaporative and condensive processes.

\section{Acknowledgements}

The generous help of Neil Cornish, Gary Gibbons, Rufus Hamade, Stephen
Hawking, Christopher Hunter, Stuart Rankin and John Stewart is
gratefully acknowledged. A.S. acknowledges the receipt of PPARC grant
number GR/L21488 and support from the DOE and NASA grant NAG 5-2788 at
Fermilab.

\section{References}

\begin{flushleft}

[1]  Teitelboim, C., (1977) Phys.Lett.B69, 240

[2]  DeWitt, B.S., (1967) Phys.Rev.160, 1113

[3]  Hartle, J.B., and Hawking, S.W., (1983) Phys.Rev.D28, 2960

[4]  D'Eath, P.D., (1996) Supersymmetric Quantum Cosmology
(Cambridge: Cambridge University Press)

[5]  Garabedian, P.R., (1964) Partial Differential Equations (New
York: Wiley)

[6]  Feynman, R.P., and Hibbs, A.R., (1965) Quantum Mechanics and
Path Integrals (New York: McGraw-Hill)

[7]  Hawking, S.W., (1979) in General Relativity.  An Einstein
Centenary Survey, ed. Hawking, S.W. and Israel, W. (Cambridge:
Cambridge University Press)

[8]  Reula, O., (1987) Max-Planck-Institut fur Astrophysik preprint
MPA 275

[9]  Hartle, J.B., and Hawking, S.W. (1976) Phys.Rev.D13, 2188

[10] Moss, I.G. and Poletti, S.J., (1994) Phys.Lett.B333, 326

[11] Misner, C.W., Thorne, K.S. and Wheeler, J.A., (1973) Gravitation
San Francisco: Freeman)

[12] Goroff, M.H. and Sagnotti, A., (1985) Phys.Lett.B160, 81

[13] D'Eath, P.D. (1994) Phys.Lett.B321, 368

[14] Wess, J. and Bagger, J. (1992) Supersymmetry and Supergravity
(Princeton: Princeton University Press)

[15] van Nieuwenhuizen, P. (1981) Phys.Rep.68, 189

[16] Choptuik, M.W. (1993) Phys.Rev.Lett.70, 9

[17] Choptuik, M.W. (1994) in Deterministic Chaos in General
Relativity ed. Hobill, D. et. al. (New York: Plenum)

[18] Christodoulou, D. (1991) Commun. Pure Appl. Math. 44, 339

[19] Christodoulou, D. (1993) Commun. Pure Appl. Math. 46, 1131

[20] Garfinkle, D. (1994) Phys. Rev.D51, 5558

[21] Gundlach, C., Price, R. and Pullin, J. (1994) Phys. Rev.D49, 890

[22] Hamade, R.S. and Stewart, J.M. (1996) Class. Quantum Grav.13, 497

[23] Press, W.H. et. al. (1994) Numerical Recipes (Cambridge:
Cambridge University Press)

[24] Iserles, A. (1996) A First Course in the Numerical Analysis of
Differential Equations (Cambridge: Cambridge University Press)

\end{flushleft}

\end{document}